\documentclass{webofc}
\usepackage[varg]{txfonts}   
\usepackage{siunitx}
\usepackage{caption,subcaption}
\usepackage{stmaryrd}
\usepackage{trimclip}
\usepackage{scalerel}
\begin{document}
\title{Exotic Meson candidates in COMPASS data}
%
%

\author{\firstname{David} \lastname{Spülbeck}\inst{1}\fnsep\thanks{\email{david.benjamin.spulbeck@cern.ch}} \\
        for the COMPASS Collaboration}

\institute{Helmholtz-Institut für Strahlen- und Kernphysik, Universität Bonn, 53115 Bonn, Germany}

\abstract{%
One of the prime goals of the COMPASS experiment at CERN is the study of the light meson spectrum, with a particular emphasis on the search for exotic states. 
The focus of this paper is on signals of the lightest hybrid candidate $\pi_1(1600)$ with spin-exotic quantum numbers $J^{PC}=1^{-+}$ in several decay channels  
such as $\pi^-\pi^+\pi^-$, $\eta^{(\prime)}\pi^-$, $\omega\pi^-\pi^0$,  $\pi^-\pi^+\pi^-\eta$, and $K_S K_S \pi$. In addition, we highlight new results for the $K^-\pi^+\pi^-$ final state, which indicate a supernumerary state with respect to the constituent quark model with $J^{P}=0^-$. 
}
\maketitle
\section{Introduction}
\label{sec:intro}

The constituent quark model (QM) describes a meson as a $q\bar{q}^\prime$-pair, where only the two quarks contribute to the total quantum numbers $J^{PC}$. However, QCD 
predicts the existence of exotic, i.e.~non-$q\bar{q}^\prime$, mesons in the form of multiquark systems, glueballs or hybrids. The latter contain
an excited gluonic field, which contributes to the total quantum numbers of the system. 
There are two sufficient signatures for the observation of exotic mesons: (i) its quantum numbers are forbidden within the QM, e.g. $0^{--}$, $(odd)^{-+}$ and $(even)^{+-}$ (spin-exotic), or (ii) more states than predicted by the QM are observed (supernumerary).
The broad and overlapping states in the light-quark sector require large data sets and state-of-the-art partial-wave analyses using large wave sets in order to identify the quantum numbers of the contributing states. 

\section{Meson spectroscopy at COMPASS}
\label{sec:ExoticMesonCandidates}
The COMPASS experiment is dedicated to investigate the structure and dynamics of light hadrons \cite{COMPASS-2008}. In the scope of its program for light meson spectroscopy, the world's largest data sample of diffractive dissociation reactions has been recorded using a negative hadron beam at \SI{190}{GeV/\textit{c}}, consisting of mainly $\pi^-$ (96.8\%) and $K^-$ (2.4\%), impinging on a liquid hydrogen target \cite{COMPASS-hadron}. This allows for precision measurements of established resonances as well as the search for new states. Isovector resonances of the $a_J$ and $\pi_J$ families in the unflavoured sector and $K_J^{(*)}$ states in the strange sector can be accessed. 

To disentangle the underlying resonances for a selected final state, a partial-wave analysis (PWA) is performed, which can be separated into two analysis stages \cite{PWAatCOMPASS}. In the first stage, the partial-wave decomposition (PWD), the data is grouped into bins of $m_X$, the invariant mass, and $t^\prime$, the reduced four-momentum transfer to the target. For each bin the strength and relative phase of each partial wave is determined using an extended log-likelihood fit. In order to do so, the full amplitude per partial wave is separated into the decay amplitude and a transition amplitude. The decay amplitude is calculated using the isobar model, in which the decay of a resonance $X$ is described via a sequence of two-body decays, and the transition amplitudes are the fit parameters. In the second analysis stage, the resonance-model fit (RMF), the transition amplitudes are parameterized by the sum of a resonant and a non-resonant component. The former is usually approximated by either the sum of relativistic Breit-Wigner amplitudes or the K-matrix approach and depends only on the invariant mass whereas the latter takes the background-dynamics in $t^\prime$ into account. A phenomenological parametrization is used for the non-resonant component. 

\subsection{Exotics in the light unflavoured sector}
\label{sec:PiOne}
\begin{figure}[!h]
    \centering
    \includegraphics[width=4.2cm]{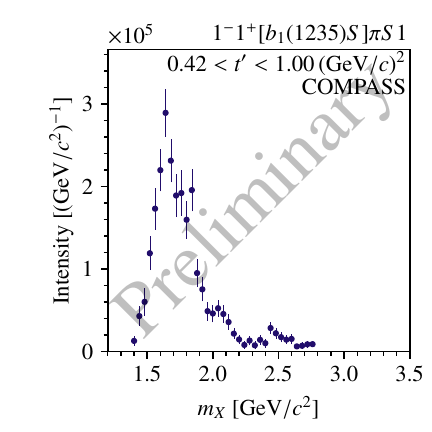}
    \includegraphics[width=4.2cm]{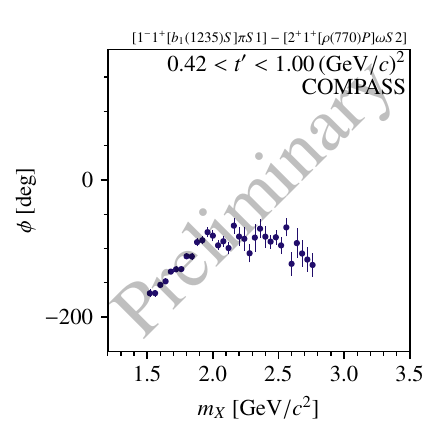}
    \includegraphics[width=4.2cm]{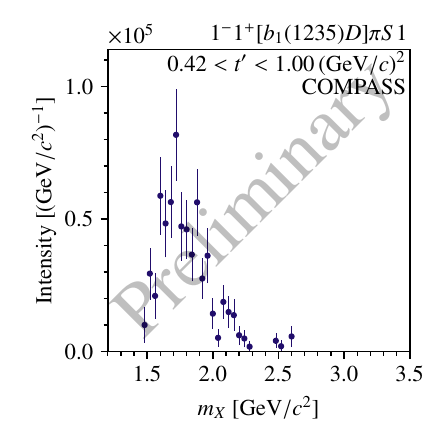}
    \qquad
    \caption{Result of the $\omega\pi^-\pi^0$ PWD in the highest $t^\prime$-bin for the $1^{-+}\,b_1(1235)\pi\,S$-wave. The intensities (phases) are sums (averages) over both charge combinations, $b_1^-\pi^0$ and $b_1^0\pi^-$. 
    The left and middle panel show the intensity and relative phase, respectively, of this wave with the isobar decay $b_1(1235)\to\omega\pi$ via an S-wave. The right panel shows the intensity of the same isobar decay via a D-wave.
    }
    \label{fig:b1pi}%
\end{figure}

\begin{figure}[h!]
\centering
\begin{minipage}[t][5.5cm][t]{.3\textwidth}
  \centering
  \includegraphics[width=3.9cm]{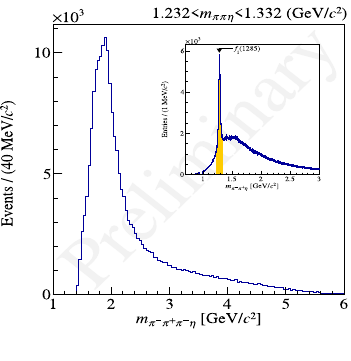}
  \captionsetup[figure]{skip=0pt}
  \captionof{figure}{Invariant mass distribution of $f_1(1285)\pi^-$ events. The insert shows the selection of $f_1(1285)$ from the $\pi^-\pi^+\eta$ subsystem.}
  \label{fig:f1pi}
\end{minipage}%
\quad
\begin{minipage}[t][5.5cm][t]{.3\textwidth}
  \includegraphics[width=3.9cm]{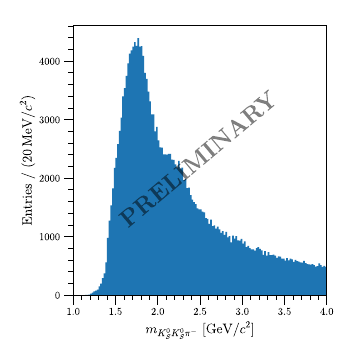}
  \captionsetup[figure]{skip=0pt}
  \captionof{figure}{Invariant mass distribution of exclusive $K_SK_S\pi^-$ events in the resonance-mass region.}
  \label{fig:KK}
  
\end{minipage}
\quad
\begin{minipage}[t][5.5cm][t]{.3\textwidth}
  \centering
  \includegraphics[width=3.9cm]{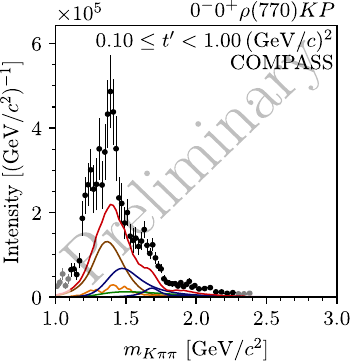}
  \captionsetup[figure]{skip=0pt}
  \captionof{figure}{Intensity (data points) and RMF (lines): resonances (brown and blue (2x)), background (orange and~green),~full model (red).}
  \label{fig:Kpipi}
\end{minipage}
\end{figure}

Several models predict the lightest hybrid mesons to have  quantum numbers $(0,\textbf{1},2)^{-+}$ or $1^{--}$ \cite{HybridPred}, which includes one spin-exotic configuration known as the $\pi_1$. Since signals in this $1^{-+}$ sector have been observed in several channels and experiments, recent lattice QCD simulations focused on the decay-channels of the $\pi_1$. The result was a dominant branching to $b_1(1235)\pi$ and comparatively suppressed branchings to $f_1(1285)\pi$, $\rho\pi$, $\eta^{(\prime)}\pi$, $f_1(1420)\pi$ and $K^*\bar{K}$. With COMPASS data the $\pi_1$ and its decay via these channels is currently being studied.
\bigbreak
$\pmb{\rho(770)\pi}$ - The flagship is the $\pi^-\pi^+\pi^-$ final state. Based on \SI{46}{M} selected events the full PWA has been performed in 11 $t^\prime$-bins taking 88 waves into account \cite{CompassRhoPiResModFit}. The analysis of the $\pi^-\pi^+\pi^-$ COMPASS data allowed for three important findings regarding the $\pi_1(1600)$-signal in the $\rho(770)\pi$-channel.
\begin{figure}[!b]
    \centering
    \includegraphics[width=4cm]{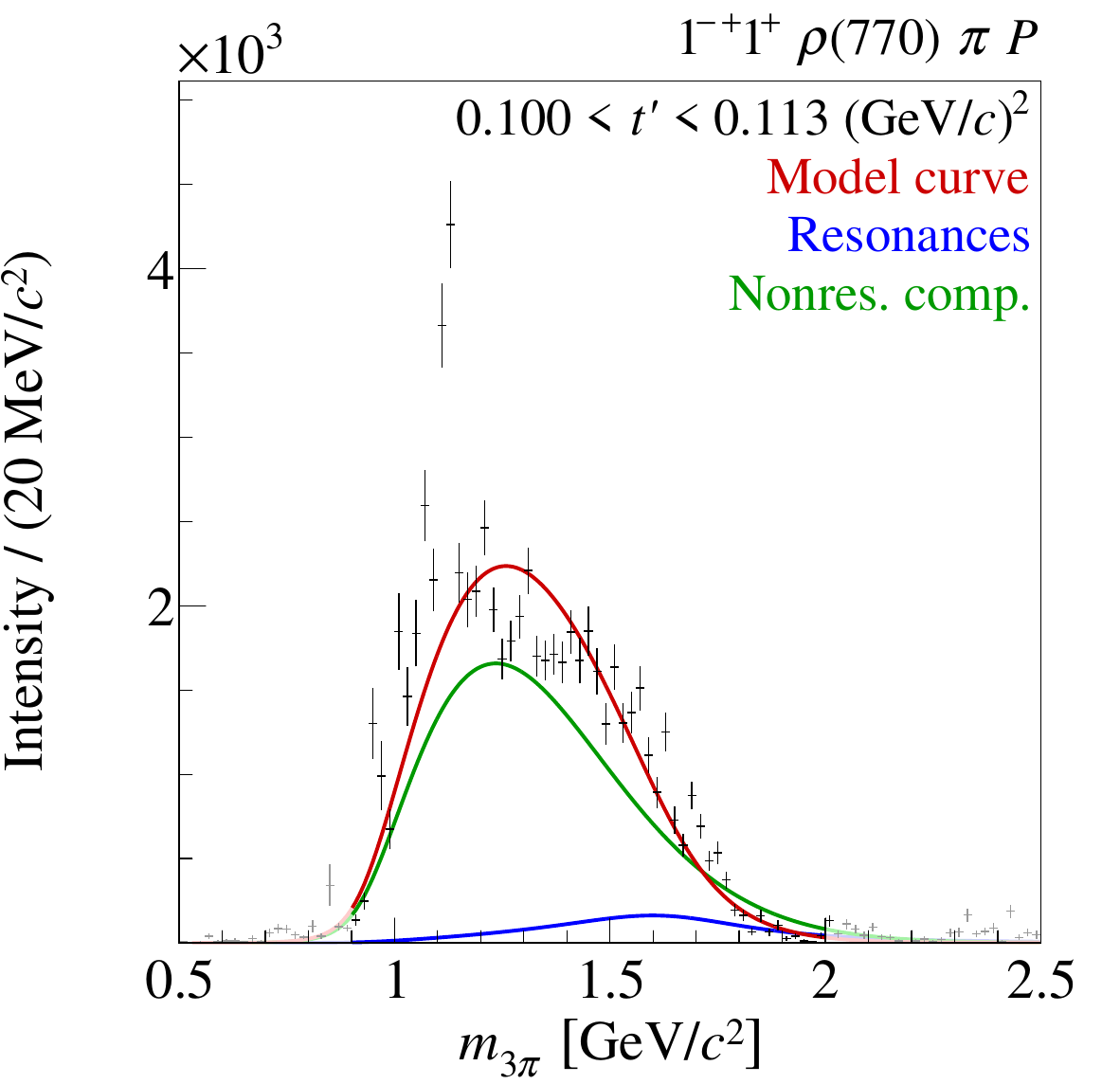}
    \includegraphics[width=4cm]{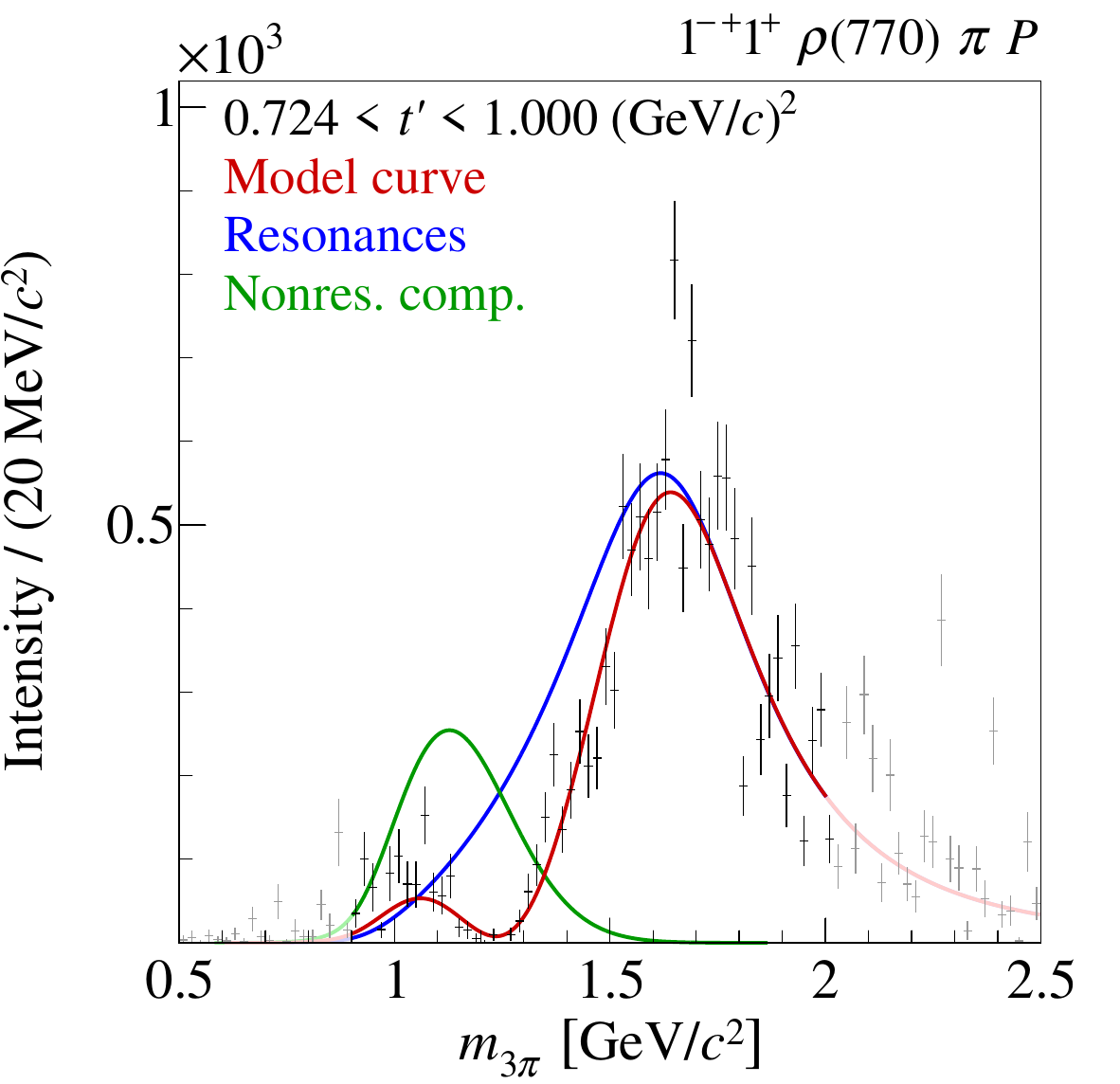}
    \includegraphics[width=4cm,clip, trim = 0 8 0 0 ]{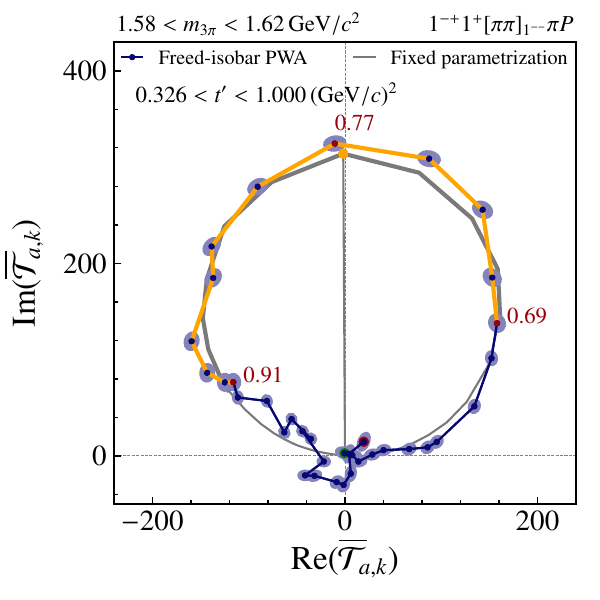}\\
    \qquad
    \caption{PWA result of the $1^{-+}1^+\rho(770)\pi^-P$-wave: Main analysis and RMF in the lowest (left) and highest (middle) $t^\prime$-bins \cite{CompassRhoPiResModFit}, Argand diagram  (right) showing an exemplary result of the freed-isobar ($\rho$) PWA \cite{CompassRhoPiFreed}.} 
    \label{fig:RhoPi}%
\end{figure}
Firstly, by covering a wide $t^\prime$-range and using a much larger wave set, COMPASS could reconcile apparent contradictions of previous experiments as analysis artefacts \cite{CompassRhoPiFreed}. 
Secondly, by splitting the data into 11 $t^\prime$-bins, a 
separation of resonant and non-resonant components in this particular wave was possible, which is necessary for a reliable extraction of the resonance parameters. From a Breit-Wigner model, a mass of $m_{\pi_1}=1600^{+110}_{-60}$ \SI{}{MeV/\textit{c}^2} (sys) and a width of $\Gamma_{\pi_1}=590^{+100}_{-230}$ \SI{}{MeV/\textit{c}^2} (sys) was extracted. The result of the RMF for the lowest and highest $t^\prime$-bin is shown in Fig.~\ref{fig:RhoPi} (left and middle, respectively) for spin projection quantum numbers $M^\epsilon=1^+$ \cite{CompassRhoPiResModFit}. 
Thirdly, a freed-isobar PWA was performed for this spin-exotic wave by removing the isobar ($\rho$) line-shape from the decay amplitude and binning the isobar mass as well. For each $m_{2\pi}$-bin a complex fit parameter was introduced accounting for the dynamics of the isobar. The result for one ($m_{3\pi}$,$t^\prime$)-bin is shown in form of the Argand diagram in Fig.~\ref{fig:RhoPi} (right) where three values of $m_{2\pi}$ are given in \SI{}{GeV/\textit{c}^2}. It shows the clear signature for the decay via the $\rho(770)\pi$ channel and by that the validity of the isobar model is demonstrated.
\bigbreak
$\pmb{\eta^{(\prime)}\pi}$ - These two channels have been analysed by other experiments before with a surprising result: the presence of two spin-exotic states with rather close masses: the $\pi_1(1400)$, observed only in the $\eta\pi^-$-channel, and the $\pi_1(1600)$, observed only in the $\eta^\prime\pi^-$-channel. 
At COMPASS these two channels are measured via the decays $\eta^{(\prime)}\hspace{-0.05cm}\shortrightarrow \pi^-\pi^+\pi^0(\eta)$ and $\pi^0(\eta)\hspace{-0.05cm}\shortrightarrow\gamma\gamma$. The PWD was performed on \SI{115}{k} ($\eta\pi^-$) and \SI{40}{k} ($\eta^\prime\pi^-$) events without any $t^\prime$-binning \cite{COMPASSEtaPrimePi}. The RMF of the COMPASS data, performed in \cite{JPAC} as a coupled channel fit using the K-Matrix approach, 
requires only one pole to describe the data. An analysis of Crystal Barrel data confirmed this finding \cite{Kopf}. The resulting resonance parameters ($m_{\pi_1}=1564\pm24\pm86$ \SI{}{MeV/\textit{c}^2} and $\Gamma_{\pi_1}=492\pm54\pm102$ \SI{}{MeV/\textit{c}^2} \cite{JPAC}) are compatible with the ones of the $\pi_1(1600)$ extracted by COMPASS for the $\pi^-\pi^+\pi^-$ final state. 
Currently the analysis is redone including the full data set, which doubles the number of exclusive events and allows us to take into account the $t^\prime$-dynamics. In addition, we will extend the mass range to higher masses, allowing us to apply constraints from Regge theory \cite{JPAC-regge} in the analysis. 
\bigbreak
$\pmb{b_1(1235)\pi}$ - Based on much smaller data sets, previous experiments observed a clear signal in the spin-exotic sector for this channel. In \cite{BNLb1} even a second state at higher masses ($\sim$\SI{2.0}{GeV/\textit{c}^2}) was suggested which still needs confirmation.
At COMPASS this channel is studied by performing the PWA on the $\omega\pi^-\pi^0$ system. Experimentally the $\omega$ is detected via the decay into $\pi^+\pi^-\pi^0$. Based on \SI{720}{k} exclusive events, which are grouped into four $t^\prime$-bins, the PWD has been performed resulting in clear signals in the spin-exotic sector and in the expected mass region of the $\pi_1(1600)$. For the highest $t^\prime$-bin the $b_1\pi$ S-wave intensity and relative phase with respect to the $\rho\omega$ D-wave are shown in Fig.~\ref{fig:b1pi} (left) and (middle) for the isobar decay $b_1\hspace{-0.05cm}\shortrightarrow\omega\pi$ via S-wave. For the same isobar but decay via D-wave, the $b_1\pi$ S-wave intensity is shown in Fig.~\ref{fig:b1pi} (right). 
A RMF is currently being performed in order to extract resonance parameters. 
\bigbreak $\pmb{f_1(1285)\pi^-} \&\pmb{K^*\bar{K}}$ - The $f_1(1285)\pi^-$ channel is analysed at COMPASS via a four-body PWA of the $\pi^-\pi^+\pi^-\eta$ system based on \SI{620}{k} events. For the events with at least one $\pi^-\pi^+\eta$ combination inside the $f_1(1285)$-range, the invariant mass is plotted in Fig.~\ref{fig:f1pi} showing a broad and peaking structure in the resonant region. 
The $K^*\bar{K}$ channel is analysed by performing a PWA of the $K_SK_S\pi^-$ system where the two $K_S$ are experimentally detected via the decay into $\pi^+\pi^-$. We have selected \SI{240}{k} exclusive events. The invariant mass distribution is shown in Fig.~\ref{fig:KK}. 
Conclusions on the presence of the $\pi_1(1600)$ shall be drawn based on the results of the PWA, which is currently work in progress for both systems.

\subsection{Exotics in the strange sector}
Using the $K^-$ component of the beam, the light strange sector can be accessed as well. Our new analysis of the $K^-\pi^+\pi^-$ final state identifies eight resonance-like signals. The most interesting one is compatible with the not PDG-established $K(1630)$ and has $J^P=0^-$. In this sector only two excited states are predicted by the QM below \SI{2}{GeV/\textit{c}^2}, but three states are observed experimentally. The lightest $K(1460)$ and the heaviest $K(1830)$ can be assigned to ordinary $q\bar{q}^\prime$ states, which makes the $K(1630)$ a supernumerary candidate. In order to achieve a stable fit for the $0^-\,\rho(770)K\,P$-wave, which is the only robust one in this sector, the parameters of the well-known $K(1460)$ were fixed to the PDG values. In a systematic study the third resonance was removed from the RMF, resulting in a worse $\chi^2_{\text{red.}}$. Translating this into a significance yields a value of $8.3\sigma$ for the presence of three states. The intensity distribution as well as the result of the RMF is shown in Fig.~\ref{fig:Kpipi} for the $0^-\,\rho(770) K\,P$-wave in the lowest $t^\prime$-bin.
\label{sec:StrangeSector}
\section{Conclusions}
Recording the world's largest data set of diffractive dissociation reactions with a $\pi^-$ beam, COMPASS entered the era of precision spectroscopy of light mesons. Including all data recorded allows us to include large wave sets and to perform a binning in both the invariant mass and squared momentum transfer. 
We report on new investigations on the spin-exotic $\pi_1(1600)$ in different decay channels. The decay branching will be crucial for the model interpretation of this state. Using data taken with a $K^-$ beam, we perform a first PWA of the strange meson spectrum in COMPASS and find evidence for a supernumerary state with $0^-$ quantum numbers. 


%

\begin{thebibliography}{}
%
%

\bibitem{COMPASS-2008}
P. Abbon et al. (COMPASS Collaboration), Nucl. Instrum. Meth. A \textbf{577} 455-518 (2007)
\bibitem{COMPASS-hadron}
P. Abbon et al. (COMPASS Collaboration), Nucl. Instrum. Meth. A \textbf{779}, 69 (2015)
\bibitem{PWAatCOMPASS}
B. Ketzer et al., Progress in Particle and Nuclear Physics, Volume \textbf{113}, 103755 (2020)
\bibitem{HybridPred}
B. Ketzer, Proceedings of Science, Volume \textbf{157}, 025 (2012) 
\bibitem{LQCD}
Antoni J. Woss et al., Phys. Rev. D \textbf{103}, 054502 (2021)
\bibitem{CompassRhoPiResModFit}
M. Aghasyan et al. (COMPASS Collaboration), Phys. Rev. D \textbf{98}, 092003 (2018) 
\bibitem{CompassRhoPiFreed}
G. D. Alexeev et al. (COMPASS Collaboration), Phys. Rev. D \textbf{105}, 012005 (2022) 
\bibitem{COMPASSEtaPrimePi}
C. Adolph et al. (COMPASS Collaboration), Phys. Lett. B, Volume \textbf{740}, 303-311 (2015) 
\bibitem{JPAC}
A. Rodas et al., Phys. Rev. Lett. \textbf{122}, 042002 (2019) 
\bibitem{Kopf}
B. Kopf et al., Eur. Phys. J. C \textbf{81}, 1056 (2021) 
\bibitem{JPAC-regge}
Ł. Bibrzycki et al. (JPAC Collaboration), Eur. Phys. J. C \textbf{81},  647 (2021)
\bibitem{BNLb1}
M. Lu et al. (E852 Collaboration), Phys. Rev. Lett. \textbf{94}, 032002 (2005) 
\bibitem{BNLf1}
 J. Kuhn et al. (E852 Collaboration), Phys. Lett. B \textbf{595}, 109-117 (2004)
\end{thebibliography}
%
%

\end{document}